# Simultaneous remote transfer of accurate timing and optical frequency over a public fiber network

Olivier Lopez[1], Amale Kanj[2], Paul-Eric Pottie[2], Daniele Rovera[2], Joseph Achkar[2], Christian Chardonnet[1], Anne Amy-Klein[1] and Giorgio Santarelli[2,3]

[1]Laboratoire de Physique des Lasers, Université Paris 13, Sorbonne Paris Cité, CNRS, 99 Av. Jean-Baptiste Clément, 93430 Villetaneuse, France
[2]LNE-SYRTE, Observatoire de Paris, CNRS, UPMC, 61 Av. de l'Observatoire, 75014 Paris, France
[3]Laboratoire Photonique, Numérique et Nanosciences, UMR 5298 Université de Bordeaux 1, Institut d'Optique and CNRS, 351 Cours de la Libération, 33405 Talence, France
Corresponding author: giorgio.santarelli@institutoptique.fr

**Abstract**
In this work we demonstrate for the first time that it is possible to transfer simultaneously an ultra-stable optical frequency and precise and accurate timing over 540 km using a public telecommunication optical fiber network with Internet data. The optical phase is used to carry both the frequency information and the timestamps by modulating a very narrow optical carrier at 1.55 µm with spread spectrum signals using two-way satellite time transfer modems. The results in term of absolute time accuracy (250 ps) and long–term timing stability (20 ps) well outperform the conventional Global Navigation Satellite System or geostationary transfer methods.

## 1. Introduction

In the last five years ultra-stable optical fiber links have been successfully developed enabling precise and accurate frequency transfer with fractional frequency stability in the range of $10^{-18}$ after only 3 hours of measurement and frequency accuracy of a few $10^{-19}$. Recently ground-breaking frequency transfer has been demonstrated on a record distance of 920 km on dedicated fiber [1]. We extended this technique to public fiber networks with simultaneous data traffic, providing a scalable technique for extension to the continental level [2,3]. This approach is much more stable and accurate than satellite-based frequency comparisons using the Global Navigation Satellite System (GNSS), or other satellite-based techniques. This opens the way to accurate remote clock comparisons, for modern atomic clocks having already demonstrated accuracy in the range $10^{-16}$-$10^{-17}$ [4-8]. This is a key-point in advanced time-frequency metrology and for advanced tests of fundamental physics [9]. Accurate frequency plays also a key role for geodesy, high resolution radio-astronomy, modern particle physics, and for the underpinning of the accuracy of almost every type of precision measurement. For all these applications, accurate timing is also important and gives the capability to precisely synchronize distant experiments. A salient case is the neutrinos speed measurement from CERN to Gran Sasso. Fiber-optical two-way time transfer methods have been demonstrated on dedicated links with an accuracy of the order of one hundred of ps or better [10,11]. Long distance accurate time dissemination is usually based on GNSS signals, or geostationary telecommunication satellites, with timing accuracy of the order of 1 ns in the best case [12,13]. In this work we present a novel method to simultaneously disseminate an ultra-stable optical frequency and accurate timing over a public telecommunication network on a 540 km optical link simultaneously carrying Internet data traffic, using a dedicated "dark" channel.

## 2. Experimental set –up description.

This method exploits an ultra-stable laser at 1.55 µm to carry both the frequency information as shown in Ref [3] and the timing signal through optical phase modulation. Figure 1 shows the scheme of the experiment. The 540 km-long optical link (LPL-Reims-LPL) depicted in Fig. 1 (i) is identical to the one reported in [3]. The link starts and ends at the LPL laboratory (Laboratoire de Physique des Lasers, Université Paris 13) in order to have the possibility to compare the signals at both ends and evaluate the distribution performance. It uses the fibers of the French National Research and Education Network (NREN) RENATER. It is composed of five different fiber spans. In each span, there are two identical parallel fibers. The third, fourth and fifth spans are long-haul intercity links simultaneously carrying internet data traffic. Optical Add-Drop

Multiplexers (OADMs) enable to extract and insert the science signal into the telecommunication fibers. Total end-to-end attenuation for the 540 km link is in excess of 165 dB. With the help of six bidirectional EDFAs and a total amplification of about 100 dB, the net optical losses exceed 65 dB.

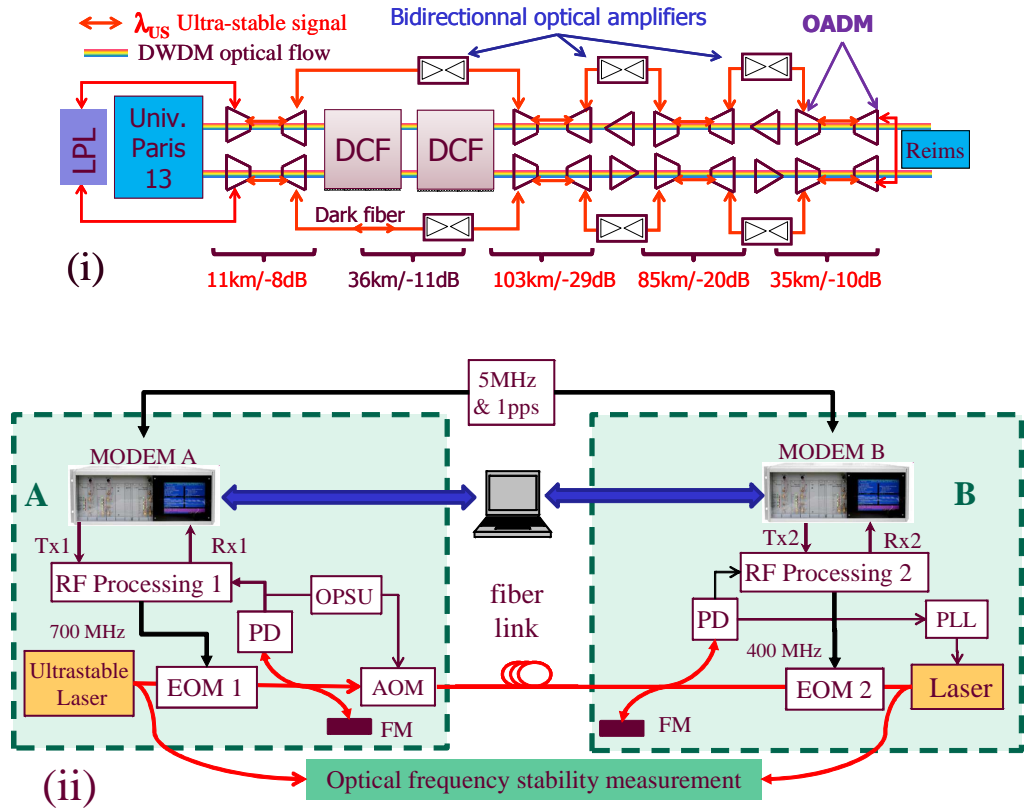

Figure 1 (i) layout of the long distance optical fiber link, OADM (Optical Add-Drop Module), DWDM (Dense Wavelength Division Multiplexing), DCF (Data Center facility) (ii) Sketch of the experimental set-up; FM : Faraday mirror, PD : photodiode, AOM : acousto-optic modulator, EOM : electro-optic modulator, OPSU : Optical Phase Stabilization Unit, PLL : Phase-Locked Loop.

Figure 1 (ii) shows the hybrid time comparison and ultra-stable optical frequency distribution system. The detailed description of the ultra-stable optical frequency dissemination is given in [3].

Here we briefly recall the operation of the optical ultra-stable frequency transfer: the frequency signal consists of the frequency of an ultra-stable cavity-stabilised laser (in block A, Fig. 1(ii)), which feeds the optical fiber. The fiber propagation noise is detected and compensated by comparison of the input optical phase with the signal phase after a round-trip, thanks to an optical interferometer. Corrections are applied using an Acousto-Optical Modulator (AOM) operating around 40 MHz. For the sake of simplicity we gather all error signal processing functions into a symbolic box called Optical Phase Stabilization Unit (OPSU). At the far end of the link the optical signal is regenerated by a narrow linewidth laser (3 kHz) (in block B, Fig. 1(ii)), phase-locked on the incoming light, and frequency-offset by 79 MHz. The optical frequency transfer stability is obtained by measuring the optical beat-note between the ultra-stable laser from A and the phase locked laser in part B.

Concerning time transfer, signals are provided by a pair of two-way satellite time transfer modems [15]. These signals are generated by relating the phase of a pseudorandom noise modulation (20 Mchip/s) on a radio frequency carrier signal (50-80 MHz) to the one-pulse-per-second (1 pps) and 5 MHz reference signals from a common clock. Orthogonal specific pseudo-random codes are allocated to each modem device. This equipment correlates the received signal with a local replica of the signal expected from the transmitting site and measures the time of arrival of the received signal with respect to the local clock. A computer collects the time of arrival of both modems and computes the differential time delays.

We implement an approach similar to the one used in coherent optical telecommunication to encode the time signal on the optical carrier. At each link end the laser is phase-modulated by a fiber pigtailed electro-optical modulator (EOM) with frequency shifted replicas of the time signal at 400 MHz and 700 MHz respectively. These frequency shifts are chosen to avoid interference between optical signals and allow efficient filtering of the time comparison signals. The low modulation depth (~1%) and the spread spectrum nature of the modulating time signals lead to a very pure optical spectrum, allowing the optical frequency carrier transfer to operate without degradation. The time signals are recovered by an optical heterodyne beat-note of the local laser with the incoming signal at each link end. After successive frequency mixing and filtering the spread spectrum time signals are processed by the modems. This step is not trivial considering the frequency width of the pseudo-random codes (~20 MHz) and the large dynamic range of the signals in the detection system, where parasitic signals due to stray reflections are 40 dB larger than the useful signal.

## 3. Results and discussion.

The frequency transfer stability and the timing stability/jitter were simultaneously measured and are plotted in Fig. 2. The frequency stability of the link reaches a resolution of $2 \times 10^{-18}$ at 30 000 s averaging time, which is almost identical to the one reported in [3]. The accuracy of the frequency transfer is about $2 \times 10^{-18}$.

The time stability shows a noise of less than 20 ps over the whole measurement time period. Delay calibration is a prerequisite of accurate dissemination of timing signals. Since the fiber optical length is unknown and fluctuates over time, a rigorous calibration test is needed to guarantee the independence of the system from the fiber length. This calibration procedure is a standard method in time&frequency metrology [15]. We change the propagation length by "shortcuts" in the accessible places along the long-haul optical link. We vary the link length from 10 m to 94 km, 400 km and the total length of 540 km. The overall link attenuation is maintained nearly constant within ±2 dB by inserting a variable optical attenuator into the common fiber path.

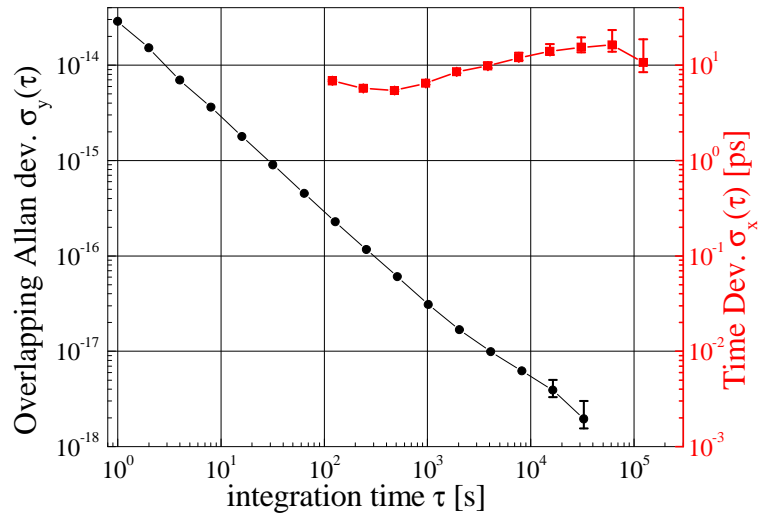

Figure 2 fractional optical frequency stability (black circles, left scale) and timing stability in picoseconds (red squares, right scale) for the 540-km fiber link.

This minimizes the impact of the received power dependent delay variations in the modems. We measure that the differential time delay variation versus distance is at most 50 ps. In addition we checked the system sensitivity by changing the power of the signals from the optical detection up to the modem input. As long as the modems operate in optimal conditions (i.e. low input power level), the system shows a coefficient below 15ps/dB. We also performed several tests of disconnection/connection, power shutdown and restart without measuring appreciable time delay variation. Tests on fiber spools of 25 km, 50 km, 75 km and 100 km are consistent with the above reported results. This method is strongly immune against fiber chromatic dispersion due to the very low frequency/wavelength difference between the two counter-propagating time signals (about 0.3 GHz/2.4 pm). The overall effect for the 540 km link is below 25 ps. The timing fluctuations due to Polarisation Mode Dispersion (PMD) penalty is also well below 20 ps, derived from previous telecommunication fiber network characterization.

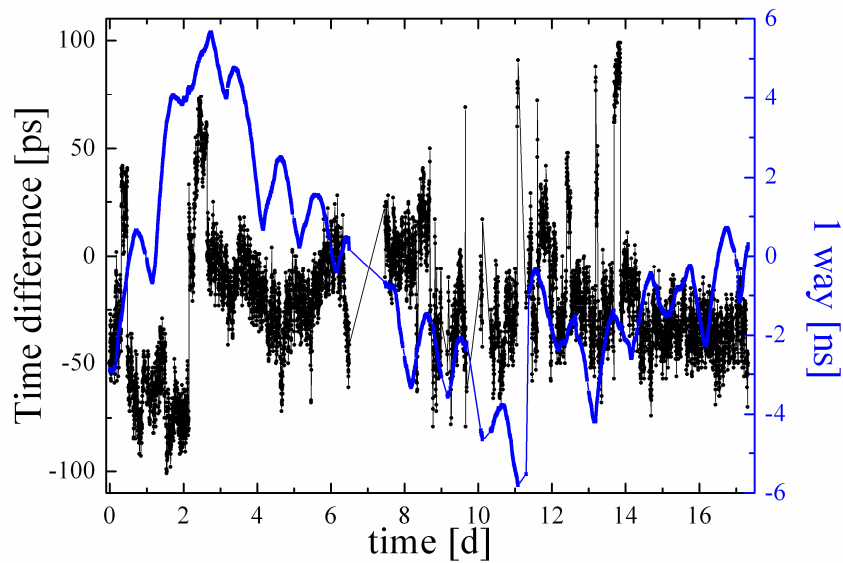

Figure 3: One-way time delay (blue plot, right axis) and differential delay (black plot, left axis) (each data point is the average of 120 one second delay measurements) versus time for the 540-km fiber link (constant calibration time offset removed on both plots).

In addition, we varied the input polarization state using Lefevre's three-loop rotating wave plate and did not observe any significant variation within a measurement resolution of 50 ps. It is worth noting that both time measurements are collocated in this experiment so that the set-up is not sensitive to the Sagnac effect. The preliminary conservative accuracy budget of 250 ps is mainly dominated by phase jumps of about 50-80 ps. The Figure 3 shows typical data over a period of more than 2 weeks (missing data are due to technical issues). The origin of such jumps seems related to the propagation in the long-haul Internet fiber link. Additional tests performed by replacing the long-haul link with fiber spools did not reveal any phase jumps over several days of measurements. The cause of such behavior is not yet well understood. Its analysis is difficult as these scarce phase jumps happen randomly. Nevertheless the system is quite robust, in addition to the results presented in Figure 3, we have several runs of about one week for which the peak-to-peak time fluctuations are below 200 ps while one-way fluctuations are in excess of 10 ns. The results in terms of timing stability and accuracy clearly outperform the satellite techniques [12].

We believe that in the case of slightly lower fiber losses and better distribution of the optical amplification such a link can be extended over more than a thousand kilometres. This method of accurate time transfer can be extended to segmented optical links by the use of intermediate optical regeneration stations [2,3] which include RF signal processing techniques and station delay calibration procedures. This method can be improved for example by developing new modems with wider pseudo-random codes. The scheme proposed here can be drastically simplified if one focuses only on time transfer as no ultra-stable laser is then needed. It is important to point out that the use of heterodyne detection allows operation with very large optical losses, for which classical Intensity Modulation (IM) techniques are ineffective. This method has been proven to be applicable to non-dedicated fiber and installed long-haul fiber links carrying Internet data. This opens the way to frequency and time dissemination on a continental scale since NRENs together with transnational fiber networks such as GEANT in Europe could be used for that purpose.

**Acknowledgments**

This work would not have been possible without the support of the GIP RENATER. We acknowledge the funding support from the Agence Nationale de la Recherche (ANR BLANC 2011-BS04-009-01).